\documentclass[reprint,amsmath,amssymb]{revtex4-1}
\usepackage[utf8x]{inputenc}
\usepackage{graphicx}% Include figure files
\usepackage{dcolumn}% Align table columns on decimal point
\usepackage{bm}% bold math
\usepackage{color} 
\usepackage{amsmath}% bold math
\usepackage{tensor}% bold mathpp

\usepackage{soul} % strikethrough
\usepackage[normalem]{ulem} % strikethrough

\begin{document}

\preprint{APS/123-QED}

%\title{Localization in quasi-periodic systems from calculations with periodic boundary conditions}

\title{Disorder averaging in random lattice models with periodic boundary conditions: Application to models with uncorrelated and correlated disorder}

\author{Bal\'azs Het\'enyi$^{1,2}$, Luis Miguel Martelo$^{3,4}$ and Andr\'{a}s L\'{a}szl\'{o}ffy$^{5,1}$}
\affiliation{$^1$Department of Theoretical Physics, Budapest University of Technology and
  Economics, H-1111 Budapest, Hungary \\ and \\
 $^2$Institute for Solid State Physics and Optics, HUN-REN Wigner Research Centre for Physics,  H-1525 Budapest, P. O. Box 49, Hungary \\ and \\
 $^3$ Departamento de Engenharia  Qu\'imica e Biol\'ogica \\ Area de Engenharia F\'isica, Faculdade de Engenharia, \\
 Universidade do Porto,  Rua Dr.  Robert Frias, 4200-465 Porto, Portugal, \\ and \\
$^4$Centro do F\'isica do Porto, Rua do Campo Alegre 687,   4169-007 Porto, Portugal, \\ and \\
$^5$ Faculty of Information Technology and Bionics, Pázmány Péter Catholic University, 1083 Práter u. 50/A, Budapest, Hungary}

\date{\today}% It is always \today, today,

\begin{abstract}
Periodic boundary conditions are not always used in the study of disordered systems, but it can be advantageous to apply them to mimick thermodynamically large systems.  In this case, polarization and its cumulants can not be obtained directly, but through the tools of the modern theory of polarization.  This theory casts the polarization in crystalline systems as a geometric phase, rather than an operator expectation value.  We develop disorder averaging techniques within the context of this theory which can calculate the variance of the polarization, its higher order moments, and the excess kurtosis (or Binder cumulant).   We also derive an indicator of delocalization based on the degeneracy as a function of boundary conditions.   We apply the computational techniques to two model systems.  To test localization, we use a one-dimensional disordered model which is fully Anderson localized.  Our calculations verify this.  We also apply our techniques to the one dimensional de Moura-Lyra model, developed to study power law correlated (controlled by a parameter, $\alpha$) disorder.   While this model is a pathological one, our method is validated.  We also point out the significance of pairwise degeneracies found in the parameter range, $\alpha>2$ and near the band center (or near half filling), where the model was conjectured to exhibit a mobility edge.
\end{abstract}


\maketitle

\section{Introduction} 

Localization of particles is a fundamental phenomenon in physics~\cite{Kramer93}.  The distinction between conductors and insulators was thought for a long time to be the result of single charge carrier localization.   The advent of quantum mechanics changed this view: the seminal work of Walter Kohn showed~\cite{Kohn64} that "many-body localization"~\cite{MBLnote} determines the difference between conductors and insulators.   The quantitative testing of this tenet in the case of bulk crystalline systems had to wait a few decades, until the development of the modern polarization theory (MPT).  In this theory, the bulk polarization of a crystalline system is cast~\cite{King-Smith93,Resta94,Vanderbilt18,Resta00,Resta98,Spaldin12} as a geometric phase~\cite{Berry84,Zak89}.  The formalism provides for the calculation of the variance of the polarization~\cite{Resta99} as well as sum rules based on higher order cumulants~\cite{Souza00}.\\

The question of localization turns out to be particularly important in disordered systems~\cite{Langedijk09,Evers08,Lee85}.  The one parameter scaling theory of Anderson and colleagues revealed~\cite{Anderson58,Abrahams79} that the nature of localization in systems with uncorrelated disorder is highly dependent on dimensionality leading to complete localization in one, crossover in two, and a mobility edge in three dimensions.   Although this fundamental paradigm was established long ago, the question of disorder effects is of perennial interest, since the discovery of almost any new physical system immediately raises related issues.  Some of the most recent examples of such studies are: structural, rather than on-site, disorder effects~\cite{Cheng24,Bhattacharjee25}, disorder effects on hyperbolic lattices~\cite{Li24}, on ring nanojunctions~\cite{Mondal25}, or the interplay between topology~\cite{Bernevig13,Asboth16} and disorder~\cite{Li09,Jiang09,Prodan11,Ryu12,Wu16,Orth16,Meier18,Awoga24,Ahmed25}.    The theoretical techniques used to gauge localization in disordered systems include calculation of the conductivity~\cite{Abrahams79} inverse participation ratio~\cite{Kramer93} (and extensions), the localization length~\cite{Kramer93}, however, there are relatively few instances~\cite{Bendazzoli10,Varma15,Hetenyi21} of the application of MPT to disordered systems.\\

Disordered systems exhibit several unique characteristics.   The three dimensional Anderson model was the first model to exhibit a mobility edge, an energy level separating conducting and insulating single particle states.  Interacting disordered systems can also exhibit the phenomenon of many-body localization~\cite{Basko06,Nandkishore15} (MBL).  MBL refers to non-ergodic behavior in such systems accompanied by an absence of thermalization.   MBL systems can exhibit a many-body mobility edge, an energy level separating delocalized and localized systems.  Li et al. investigated~\cite{Li16} three models which exhibit single particle mobility edges (two extended Aubry-Andr\'{e} models and the three dimensional Anderson model).  The main aim of this study was the investigation of many-particle mobility edges for models which exhibit single particle ones.  A central finding of this study was the existence of a phase which is non-ergodic but metallic.

Li et al.~\cite{Li16} calculated the entanglement entropy and subsystem particle number fluctuations.  Additional tools used in MBL studies include entanglement spectra, level repulsion  statistics and the inverse participation ratio, although, the meaning of the latter is not clear in many-body systems.   It is natural to ask whether MPT is connected to MBL.  Ryu and Hatsugai~\cite{Ryu06} used a modified twist operator to calculate the entanglement entropy in the Su-Schrieffer-Heeger model.  A recent study by Faugno and Ozawa investigated~\cite{Faugno26} the use of MPT localization length and quantum geometric quantities to gauge MBL. 

In the model originally studied by Anderson, disorder was uncorrelated.  In real materials, however, disorder tends~\cite{Simonov20,Wei24,Goodwin25} to be correlated.   This happens to be the case for simple metallic systems~\cite{Dunlap90}, metal-organic frameworks~\cite{Meekel21}, superconductors~\cite{Neverov22,Neverov25} or DNA strands~\cite{Peng92}.    Recently  efforts~\cite{Simonov20,Li25} focused on synthesizing materials with controlled correlated disorder, a development essential for applications, such as  supercapacitors~\cite{Liu24}.  Correlated disorder has also been studied extensively from a  theoretical point of view.   One attempt to capture power law correlated disorder is the de Moura-Lyra model~\cite{deMoura98} (dMLM).  The idea of this model is to generate a self-affine random potential exhibiting power law decay via the inverse Fourier transform method.   The model parameter $\alpha$ controls the exponent of the Fourier components.  The dMLM already has a long history~\cite{deMoura98,Dominguez-Adame03,Kantelhardt00,Bunde00,Russ01,Shima04,Kaya07,deMoura00,Izrailev12,Petersen13,SantosPires19,Duthie22,Khan22,Khan23}.    One motivation for its study was that initial investigations reported the presence of a mobility edge~\cite{deMoura98,Dominguez-Adame03} (for $\alpha>2$ the states near the center of the band extended states were found).   Other studies pointed to difficulties with the thermodynamic limit~\cite{Kantelhardt00,deMoura00,Bunde00,Russ01} and some numerical investigations were contradictory~\cite{Shima04,Kaya07}.   We mention two rigorous studies pointing out, specifically, some of the pathologies of dMLM.  Petersen and Sandler~\cite{Petersen13} found that the potential exhibits a size dependent anti-correlation (for $\alpha>1$).  For very large $\alpha$ the disorder potential becomes a sinusoidal superpotential, although, at short scales, randomness still persists.  Santos Pires {\it et al.}~\cite{SantosPires19} showed that a short range (single bond) uncorrelation persists even for large system sizes.    The localization length shows a global delocalization transition at $\alpha \approx 1$.  It was concluded, that, since for $\alpha>1$ all states are extended, no mobility edge exists. \\

In this work we apply disorder averaging techniques in the context of the MPT.  In particular, we calculate disorder averaged versions of the variance, its size scaling exponent, the geometric Binder cumulant.  We also develop an indicator for delocalization, which takes advantage of a feature present in finite one dimensional lattice systems: when applying a gauge field at the boundaries (equivalent to switching the system between periodic and antiperiodic boundary conditions), gapped systems can be distinguished from closed gap (degenerate) systems by calculating the expectation value of the momentum shift operator. 
We apply our methodology to two types of one-dimensional disordered: the Anderson model and the dMLM.  In addition to the usual energy dependent averaging of single particle states, we also study the particle density dependence of localization in many-body versions of both models.  Our many-body calculations for both models are motivated by Kohn's tenet~\cite{Kohn64}: ultimately, it is the many-body system and its localization that has to be investigated, rather than the single particle ones to assess charge transport properties.  Fermions obey the Pauli exchange principle, which acts as a kind of "correlation" (short range repulsion) in many particle systems.   Single particle studies entirely miss the interplay of this correlation with disorder.  We fill this gap by explicitly considering finite particle number systems (which in the absence of  interparticle potentials can be described by Slater determinants).    \\

For the Anderson model we find that the size scaling exponent at the single particle level is zero for localized states, and jumps to a value of two in the tight-binding limit.  For the many-body calculation, the localized states exhibit a size scaling exponent of unity, but still jump to two upon delocalization.   Although, the dMLM is a very pathological model, it is interesting to apply our newly developed disorder averaging method to it.  Our calculations for the size scaling exponent of the dMLM suggest a global delocalization transition at $\alpha \approx 1$, consistent with the findings of Petersen and Sandler~\cite{Petersen13}, and Santos Pires {\it et al.}~\cite{SantosPires19}.   Other delocalization sensitive parameters also increase (indicating delocalization), but they only reach their maximum values within the region that was considered the mobility edge in initial studies.  The region $\alpha>2$, near the band center, is distinct from the rest of the delocalized region: energy levels within that region close gaps in pairs, and when such  "two member bands" are partially filled, our delocalization sensitive quantities tend to exhibit their maximum values. \\

Our paper is organized as follows.  In the next section we present the tools of the modern polarization theory, as developed in Refs. \cite{Hetenyi19,Hetenyi22,Hetenyi24}.  After defining statistical moments and cumulants, we derive the quantities useful for calculations (the variance, the geometric Binder cumulant).  We also validate the derived expressions for some simple cases.  In section \ref{sec:dsrdr} we explain how disorder averaging can be applied.  In sections \ref{sec:Anderson} and \ref{sec:dML} we present our numerical results for a one-dimensional Anderson and de Moura-Lyra models, respectively.  In section \ref{sec:cnclsn} we conclude our work.\\

\section{Gauging localization in systems with periodic boundary conditions}

\label{sec:MPT}

In this section we assemble the tools of the modern polarization theory which are useful in gauging localization.  We first introduce the characteristic function, from which moments and cumulants can be derived for ordinary probability distributions.  In crystalline systems the relevant probability distribution is periodic, which means that the moments and cumulants are approximate, because the characteristic function only exists on a discrete set of points.  Therefore, moments and cumulants in this case corresond to approximate (finite difference) derivatives, rather than continuous ones.  Centering the distribution (by taking the absolute value of the characteristic function) and avoiding logarithmic derivatives leads to moments which preserve scaling information.  The modern polarization analog of the Binder cumulant can be constructed.  A renormalization approach based on the characteristic function is also introduced.\\

\subsection{Moments and cumulants of a probability distribution}

Given a probability density function, $P(x)$, which obeys,
\begin{equation}
P(x) \geq 0, \int_{-\infty}^\infty dx P(x) = 1.
\end{equation}
The characteristic function of $P(x)$ is its Fourier transform, defined as,
\begin{equation}
\label{eqn:fk}
f(k) = \int_{-\infty}^\infty dx P(x) \exp(i k x).
\end{equation}
The $n$th moment, $M_n$ of $P(x)$ can be obtained from $f(k)$ as,
\begin{equation}
M_n = \frac{1}{i^n} \left. \frac{\partial^n}{\partial k ^n} f(k) \right|_{k=0} = \int_{-\infty}^\infty dx P(x) x^n.
\end{equation}
The $n$th cumulant of $P(x)$ can be obtained from $f(k)$ by taking logarithmic derivatives, as,
\begin{equation}
C_n = \frac{1}{i^n} \left.  \frac{\partial^n}{\partial k ^n} \ln f(k) \right|_{k=0}.
\end{equation}
Cumulants and moments are related to each other, the first few cumulants can be written in terms of moments as,
\begin{eqnarray}
\label{eqn:cmlnts_mmnts}
C_1 &=& M_1 \\ \nonumber
C_2 &=& M_2 - M_1^2 \\ \nonumber
C_3 &=& M_3 - 3M_2 M_1 + 2 M_1^3 \\ \nonumber
C_4 &=& M_4 - 4M_3 M_1 - 3 M_2^2 + 12 M_2 M_1^2 - 6 M_1^4.
\end{eqnarray}
The first cumulant is the mean, the second cumulant is the variance (its square root is the standard deviation).  Another useful set of cumulants are the centered cumulants, obtained by a shift in $P(x) \rightarrow \tilde{P}(x) = P(x+M_1)$, resulting in a zero first moment,
\begin{equation}
\tilde{M}_1 = \int_{-\infty}^\infty dx \tilde{P}(x) x = 0.
\end{equation}
Defining the cumulants and moments of $\tilde{P}(x)$ exactly as was done for $P(x)$ results in the centered moments and cumulants, which are related to each other as follows:
\begin{eqnarray}
\tilde{C}_1 &=& \tilde{M}_1 = 0 \\ \nonumber
\tilde{C}_2 &=& \tilde{M}_2 \\ \nonumber
\tilde{C}_3 &=& \tilde{M}_3  \\ \nonumber
\tilde{C}_4 &=& \tilde{M}_4 - 3 \tilde{M}_2^2.
\end{eqnarray}
In statistics standardized moments are often used in characterizing probability distributions.   In this work we will make use of the excess kurtosis, defined as,
\begin{equation}
K = \frac{\tilde{C}_4}{\tilde{C}_2 \tilde{C}_2}.
\end{equation}
More precisely, we will use a quantity known as the Binder cumulant $U_4$.  The definition of $U_4$ is,
\begin{equation}
U_4 = -\frac{1}{3} K = 1 - \frac{1}{3} \frac{\tilde{M}_4}{\tilde{M}_2 \tilde{M}_2}.
\end{equation}
This quantity was introduced into statistical physics by Binder~\cite{Binder81a,Binder81b}.   The original Binder cumulant corresponds to the excess kurtosis (time minus one third) of the order parameter.   Using the finite size scaling hypothesis~\cite{Fisher72a,Fisher72b} it can be shown that the Binder cumulant is size independent at critical points.  This is the property that makes this quantity useful in detecting critical points. \\

\subsection{Cumulants of the polarization in a crystalline system}

Consider a one-dimensional system on a lattice of $L$ sites labeled by the discrete coordinates $x=1,...,L$.  The system is assumed to contain $N$ particles.  In first quantized form, the position of each particle is labeled $x_j$, with $j=1,..,N$ and $x_j=1,..,L$.   The ground state wavefunction, assumed non-degenerate, in first quantized form can be written $\Psi(x_1,...,x_N)$.  Periodic boundary conditions are assumed, that is,
\begin{equation}
\Psi(...,x_j,...) = \Psi(...,x_j+L,...),
\end{equation}
for any $j$.\\

In the modern polarization theory, polarization is represented by a geometric phase, rather than an operator.  This state of affairs arises due to the presence of periodic boundary conditions.  The starting point for calculating cumulants in this case is the quantity sometimes called the polarization amplitude, 
\begin{equation}
Z_q = \langle \Psi | \exp \left( i \frac{2 \pi q}{L} \hat{X} \right) | \Psi \rangle,
\end{equation}
where, in second quantized form, $\hat{X} = \sum_{j=1}^L x \hat{n}_x$, where $\hat{n}_x$ denotes the density operator at $x$, and $q=0,...,L-1$.   Let us show that $Z_q$ is a (discrete) characteristic function.  $|\Psi \rangle$ denotes the ket corresponding to the  ground state many-body wave function.
We can write $Z_q$ as
\begin{equation}
Z_q = \sum_{x_1 = 1}^L...\sum_{x_N=1}^L |\Psi(x_1,...,x_N)|^2 \exp \left(  i \frac{2 \pi q}{L} X \right).
\end{equation}
In this case, since we switched to first quantization, $X = \sum_{j=1}^N x_j$.  We can define a probability distribution,
\begin{equation}
P(X) = \sum_{x_1 = 1}^L...\sum_{x_N=1}^L  |\Psi(x_1,...,x_N)|^2 \delta_{X,\sum_{j=1}^N x_j},
\end{equation}
for which it holds that $P(X+L)=P(X)$.  $Z_q$ can be written as,
\begin{equation}
Z_q = \sum_{X=1}^L P(X) \exp \left(  i \frac{2 \pi q}{L} X \right).
\end{equation}
Comparing $Z_q$ with $f(k)$ (defined in Eq. (\ref{eqn:fk})) we see that $Z_q$ is a characteristic function (one can associate $k$ in Eq. (\ref{eqn:fk}) with $2\pi q/L$).  A crucial difference between $Z_q$ and $f(k)$ is that the argument of $Z_q$ is discrete, while that of $f(k)$ is continuous.  For this reason, moments and cumulants can only be defined through finite difference (or other approximate) derivatives, not continuous ones.  For example, applying the lowest order finite difference approximation for the first two cumulants results in,
\begin{eqnarray}
\label{eqn:C2C4}
C_1 &=& \frac{L}{2\pi} \mbox{Im} \ln Z_1 \\ \nonumber
C_2 &=& -\frac{L^2}{2\pi^2} \mbox{Re} \ln Z_1.
\end{eqnarray}
Another difference is that the usual relations between cumulants and moments (Eq. (\ref{eqn:cmlnts_mmnts})) does not hold when the derivatives are only approximate. \\

Broadly speaking, our approach will be to calculate $Z_q$, take the absolute value of each $Z_q$ (which amounts to centering the underlying distribution), and then apply finite difference derivatives to obtain moments, rather than cumulants.  From these moments we can calculate the Binder cumulant ($U_4$).  Our approximate second and fourth moments take the form,
\begin{eqnarray}
\label{eqn:M2M4_}
\tilde{M}_2 &=& \frac{L^2}{2 \pi^2} (1 - |Z_1|),\\
\tilde{M}_4 &=& \frac{L^4}{8 \pi^4} \left(|Z_2| - 4 |Z_1| + 3\right), \nonumber
\end{eqnarray}
from which the Binder cumulant is defined as,
\begin{equation}
U_4 = 1 - \frac{1}{3} \frac{\tilde{M}_4}{\tilde{M}_2^2}.
\end{equation}

For future use we also define the size scaling exponent of the variance, $\gamma$, as $M_2 = A L^\gamma$.\\

\begin{figure}[t]
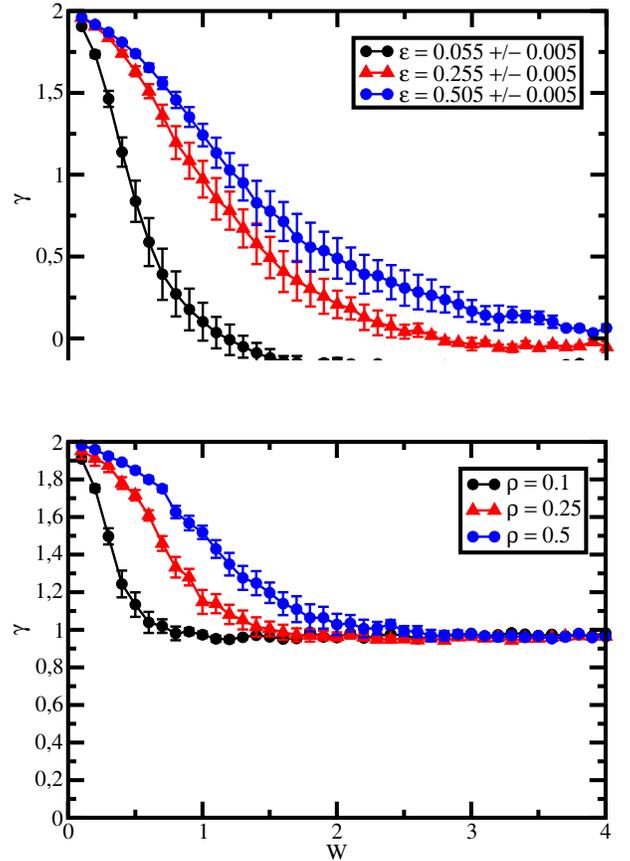

 \centering
 \includegraphics[width=8cm,keepaspectratio=true]{./gamma1D.eps}
 \includegraphics[width=8cm,keepaspectratio=true]{./gamma1D_SD.eps}
 \caption{Size scaling exponent ($\gamma$) of the second cumulant as a function of disorder strength.  The number of disorder realizations averaged was $1000$ in all calculations.  Panel (a)  shows $\gamma$ from a fit to $m_2(\epsilon) = A L^\gamma$ for three different energy intervals.  Panel (b) shows $\gamma$ from a fit to $M_2(\rho) = A L^\gamma$ for three different particle densities.    }
 \label{fig:gamma1D}
\end{figure}

\subsection{Validation}

The above methodology has been validated as a gauge for metal-insulator (or delocalization-localization) transition in a number of cases.  The essential point is that in the metallic phase (if the ground state is non-degenerate) $Z_1 \rightarrow 0$, while in the insulating phase, $Z_1 \rightarrow 1$.  The original suggestion of Resta and Sorella was to use the variance (as given in Eq. (\ref{eqn:C2C4})) as a gauge.  This expression provides quantitative results in the insulating phase, but diverges in the metallic phase (due to $\ln Z_1$ diverging), making it difficult to use.  The scaling behavior is maintained if $\tilde{M}_2$ and $\tilde{M}_4$ are taken to be of the form in Eq. (\ref{eqn:M2M4_}).  It is this feature that renders the construction of the Binder cumulant, $U_4$, possible.\\

Here, we give two simple examples which validate the above tools.  \\

Consider a tight-binding lattice model with open boundary conditions of length $L$.  The single particle wavefunctions of this system are given by,
\begin{equation}
\Psi_m(x_j) = \frac{1}{\sqrt{N_\Psi}} \sin \left( \frac{m \pi x_j}{L+1} \right),
\end{equation}
where $m$ is a quantum number, $x_j$ denotes the coordinates which range from one to $L$ and $N_\Psi$ denotes a normalization constant.   We can define the second and fourth moments as,
\begin{eqnarray}
M_2 &=& \frac{1}{N}\sum_{m=1}^N \sum_{j=1}^L x_j^2|\Psi_m(x_j)|^2 \\ \nonumber
M_4 &=& \frac{1}{N}\sum_{m=1}^N \sum_{j=1}^L x_j^4|\Psi_m(jx_)|^2 
\end{eqnarray}
We calculated $M_2$, $M_4$ for a variety of system sizes at various particle densities.   The size scaling exponent was obtained by fitting to a large set of different system sizes and we always find $\gamma=2$.   The Binder cumulant tends to the value $0.4$ in the limit of large system size.  This is due to the fact that the distribution of the total position tends to a flat distribution, for which the Binder cumulant takes this known value.\\

One can compare this to Fig. 1 of Ref. \cite{Hetenyi22} where the Binder cumulant according to moments of Eq. (\ref{eqn:M2M4_}) are calculated.  It is an easy matter to check that in the limit of $|Z_1| \rightarrow 0$ and $|Z_2| \rightarrow 0$ the Binder cumulant tends to $0.5$.  However, as the finite difference approximation is improved (Fig. 1 of Ref. \cite{Hetenyi22}) the Binder cumulant approaches a value of $0.4$. \\

On the insulating side, we quote a well-known result which relates the dielectric susceptibility ($\chi$) to the variance of the total position via second order perturbation theory.  For lattice models, this derivation was done by Baeriswyl~\cite{Baeriswyl00},
\begin{equation}
\label{eqn:sus}
\chi \leq \frac{2}{\Delta_g L} M_2,
\end{equation}
where $\Delta_g$ denotes the energy gap of the system.  In order for $\chi$ not to diverge, the size scaling exponent of $M_2$, $\gamma \leq 1$.\\

A number of calculations~\cite{Varma15,Hetenyi24} have been published which verify the above predictions.  A metal-insulator transition (delocalization-localization transition) can be determined via finite size scaling of the variance of the polarization or by calculating the geometric Binder cumulant.  The latter has the advantage that one does not have to compare different system sizes. \\

\section{Disorder averaging}

\label{sec:dsrdr}

\subsection{Disorder averaging over single particle states}

In one set of calculations we generate a large number of independent disorder realizations.  For each disorder realization we diagonalize the Hamiltonian, obtaining both energy eigenvalues and eigenstates ($E_\lambda,\phi_\lambda(j)$).  In order to be able to compare between different disorder realizations we define normalized energy eigenvalues,
\begin{equation}
\epsilon_\lambda = \frac{E_\lambda - E_n}{E_x - E_n},
\end{equation}
where $E_n$($E_x$) is the minimum(maximum) energy eigenstate within a given disorder realization.   This allows the calculation of normalized energy resolved eigenvalues.  Let us first define, for the operator $\hat{O}$, the single particle expectation value, $O_\lambda = \langle \phi_\lambda | \hat{O} | \phi_\lambda \rangle$.  Using $O_\lambda$, we define, 
\begin{equation}
O_\epsilon =  \overline{\langle O_\lambda \rangle_{\epsilon_\lambda \in [\epsilon - \Delta \epsilon,\epsilon + \Delta \epsilon] } },
\end{equation}
where $\langle \rangle_{\epsilon_\lambda \in [\epsilon - \Delta \epsilon,\epsilon + \Delta \epsilon] }$ indicates averaging over states $\lambda$ whose energy eigenvalue, $\epsilon_\lambda$ falls within the range $\epsilon - \Delta \epsilon,\epsilon + \Delta \epsilon$.  The overbar indicates averaging over disorder realizations. \\

\subsection{Finite size scaling and the geometric Binder cumulant in the analysis of single particle states}

We define the discrete cumulant generating function (in some references, known as the polarization amplitude) for a single state $\lambda$ as
\begin{equation}
Z_q^{(\lambda)} = \langle \phi_\lambda | \exp\left(i \frac{2 \pi q}{L} \hat{X} \right) | \phi_\lambda \rangle,  
\end{equation}
where $\hat{X} = \sum_{j=1}^L x_j \hat{n}_j$.   $Z_q^{(\lambda)}$ is the characteristic function of a corresponding probability distribution, $P_X^{(\lambda)}$, which can be obtained by applyint a discrete Fourier transform, as,
\begin{equation}
\label{eqn:PX}
P_X^{(\lambda)} = \frac{1}{\sqrt{L}} \sum_{q = 0}^{L-1} Z_q^{(\lambda)} \exp \left( i \frac{2 \pi q}{L}X\right).
\end{equation} 
We can center the distribution $P_X$ by taking the absolute value of $Z_q^{(\lambda)}$.   This is a necessary step in the averaging process, because we want to compare the fluctuations of the system due to the random disorder.   Centering the distributions excludes fluctuations that are due to the mere relative displacements of systems with different disorder realizations, and these fluctuations are not related to whether the system is localized or delocalized (insulating or conducting).\\

The average quantity we calculate, over disorder realizations and within energy intervals, is,
\begin{equation}
\bar{Z}_{q,\epsilon} =  \overline{\langle |Z_q^{(\lambda)}| \rangle_{\epsilon_\lambda \in [\epsilon - \Delta \epsilon,\epsilon + \Delta \epsilon] } }.
\end{equation}
Using $\bar{Z}_{q,\epsilon}$, one can define approximate second and fourth order, energy resolved, cumulants,
\begin{eqnarray}
\label{eqn:M2M4}
\tilde{m}_2(\epsilon) &=& \frac{L^2}{2 \pi^2} (1 - \bar{Z}_{1,\epsilon})\\
\tilde{m}_4(\epsilon) &=& \frac{L^4}{8 \pi^4} \left(\bar{Z}_{2,\epsilon} - 4 \bar{Z}_{1,\epsilon} + 3\right). \nonumber
\end{eqnarray}
From these approximate cumulants an energy resolved geometric Binder cumulant can be defined,
\begin{equation}
u_4(\epsilon) = 1 - \frac{1}{3}\frac{\tilde{m}_4(\epsilon)}{\tilde{m}_2(\epsilon)^2}.
\end{equation}

\begin{figure}[t]
 \centering
 \includegraphics[width=8cm,keepaspectratio=true]{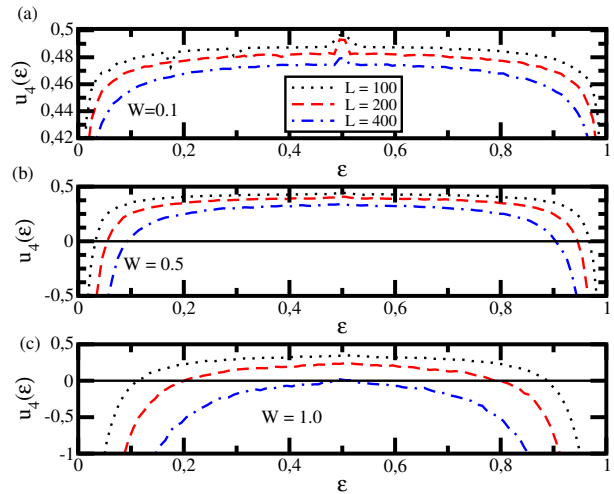}
 \caption{Geometric Binder cumulant ($u_4$) as a function of normalized energy ($\epsilon$) for a one-dimensional Anderson model for three different disorder strengths ($W = 0.1, 0.5, 1.0$) and three system sizes ($L=100, 200, 400$).  The number of disorder realizations averaged was $1000$.}
 \label{fig:U4st}
\end{figure}

\subsection{Finite size scaling and the geometric Binder cumulant in the analysis of a many-body non-interacting system}
\label{ssec:fss}

We consider a system of $N$ particles and $L$ sites (or, particle density, $\rho = N/L$).  The first step, again, is to diagonalize the Hamiltonian, and obtain single particle state, but in this case, we will form a Slater determinant of the $N$ lowest energy levels resulting in a many-body wave function,
\begin{equation}
\Psi(x_1,...,x_N) = \mbox{Det} \left[ \Phi_\lambda(x_\mu) \right]; \lambda = 1,...,N; \mu = 1,...,N.
\end{equation}
To construct the GBC, one first calculates the quantity $Z_q$, defined in this case as
\begin{equation}
Z_q(\rho)  =  \mbox{Det} \left[ U_{\lambda \lambda'}^{(q)} \right],
\end{equation}
where
\begin{equation}
U_{\lambda \lambda'}^{(q)}= \sum_{j=1}^L \phi^*_\lambda(x_j) \exp\left( i \frac{2 \pi q}{L}x_j\right) \phi_{\lambda'}(x_j).
\end{equation}

We again center the underlying distribution by taking the absolute value of all $Z_q$.   Using these centered distributions, we can define the approximate second and fourth order statistical moments as
\begin{eqnarray}
\label{eqn:M2M4}
\tilde{M}_2(\rho) &=& \frac{L^2}{2 \pi^2} (1 - |Z_1(\rho)|)\\
\tilde{M}_4(\rho) &=& \frac{L^4}{8 \pi^4} \left(|Z_2(\rho)| - 4 |Z_1(\rho)| + 3\right), \nonumber
\end{eqnarray}
From which we can define the GBC as
\begin{equation}
\label{eqn:U4}
U_4(\rho) = 1 - \frac{1}{3} \frac{\tilde{M}_4(\rho)}{\tilde{M}_2(\rho)^2}.
\end{equation}

\subsection{Exploiting the degeneracy}

It was argued by Aligia and Ortiz that $|Z_1| \rightarrow 1$ in the insulating phase and $|Z_1| \rightarrow 0$.  It is this fact that makes the quantity $Z_q$ a useful tool in distinguishing metals from insulators.  In systems of finite size this statement requires a refinement.  In Ref. \cite{Hetenyi24} it was shown that for system of finite size the many-body ground state in the delocalized (metallic) phase can be either twofold degenerate or nondegenerate.   For a detailed explanation of how this arises, see Ref. \cite{Hetenyi24}. In the case of a degenerate ground state $|Z_1| \rightarrow 1/2$.  The presence or absence of this degeneracy depends on the relative parity of the number of particles and the system size.  Also, a system with a nondegenerate ground state can be turned into a degenerate one by the application of a Peierls phase of $\Phi = \pi/L$ or vice versa.  A Peierls phase of this magnitude shifts the $k$-points representing the Brillouin zone by $\pi/L$, meaning that if the $k$-points were originally distributed symmetrically about the origin that ceases to be so and vice versa.  It is for this reason that the relative parity of the particle number and the system size determine whether the ground state is degenerate or not.  \\

Most importantly, from our point of view, in the insulating phase this degeneracy is not present, and tuning the Peierls phase makes no difference in the value of $|Z_1|$.   Although, the above state of affairs holds for finite systems, and is therefore often considered an "artifact", it can also be used as an additional tool to detect localization transitions.  We implement a Peierls phase in our many-body calculation of the dMLM by tuning the hopping parameter as,
\begin{equation}
t \rightarrow t\exp(i \Phi).
\end{equation}
We will compare two cases, one with $\Phi=0$, and one with $\Phi = \pi/L$.  We will show calculations of the difference between $|Z_1|$ for the two cases, a quantity we will name the {\it degeneracy indicator}.\\

\section{Results: Anderson model}

\label{sec:Anderson}

The model we study in this section is a simple tight-binding model with nearest neighbor hoppings and with uncorrelated on-site disorder.  Its Hamiltonian is given by
\begin{equation}
H  =  \sum_{j=1}^L \left[ -t(c_j^\dagger c_{j+1} + c_{j+1}^\dagger c_j) + W \xi_j n_j \right],
\end{equation}
where $c_j^\dagger$($c_j$) denote the creation(annihilation) operators at site $j$, $n_j = c_j^\dagger c_j$ denotes the particle density operator at site $j$, $t$ denotes the hopping parameter and $W$ denotes the disorder strength.  $\xi_j$ is a uniform random number in the range $[-\frac{1}{2},\frac{1}{2}]$.  In all of our calculations periodic boundary conditions are assumed.   \\
\begin{figure}[t]
 \centering
 \includegraphics[width=8cm,keepaspectratio=true]{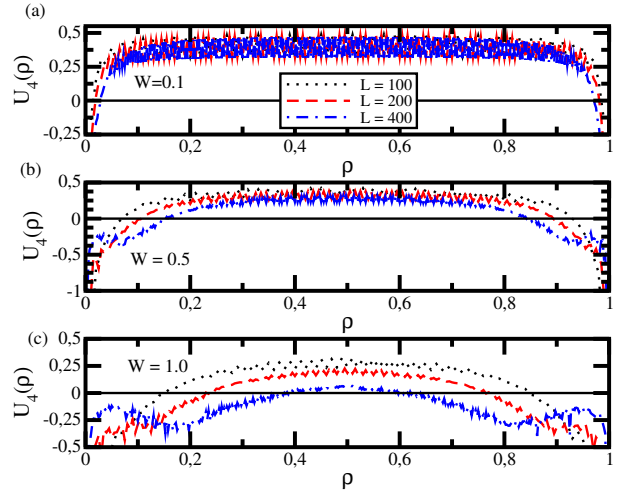}
 \caption{Geometric Binder cumulant ($U_4$) as a function of particle density ($\rho$) for a many-body one-dimensional Anderson model for three different disorder strengths ($W = 0.1, 0.5, 1.0$) and three system sizes ($L=100, 200, 400$).  The number of disorder realizations averaged was $1000$.}
 \label{fig:U4sd_1D}
\end{figure}

Fig. \ref{fig:gamma1D} shows disorder averaged calculations for the size scaling exponent of the variance, $\gamma$.  In generating these results, the variance, $M_2$, was first calculated on a grid of $50$ points in the $W$ direction, and a grid of $100$ in the $\epsilon$ direction.  The number of disorder realizations averaged was $1000$.  At each grid point a least squares fit to the function $M_2 = A L^\gamma$ was done.  Panel (a) shows results for single particle states averaged over particular energy intervals (indicated in the legend of the figure).  In the absence of disorder $\gamma=2$.  Most importantly, $\gamma$ rapidly decreases upon increasing $W$ (the disorder strength).  The system localizes very strongly, since $\gamma$ goes significantly below unity, even below zero.   The localization is stronger for energy intervals centered at values near the band edges.   Minding Eq. (\ref{eqn:sus}), we see that single particle states averaged in energy intervals are zero susceptibility states, one may say "hyperlocalized".\\

Panel (b) shows the analogous particle density dependent many-body calculations for $\gamma$.  The system sizes used were $L=100,200,400,800$.  The number of disorder realizations averaged was $1000$.  Results for three particle fillings are shown ($\rho=0.1,0.25,0.5$).  Again, $\gamma=2$ in the absence of disorder, and $\gamma$ rapidly decreases as $W$ is increased.  In this case, for large values of $W$ $\gamma \approx 1$, rather than zero, which we found for the single-particle case.  The system at half-filling is the least localized out of the three.  The tending to one of the size scaling exponent $\gamma$ leads to a finite dielectric susceptibility (Eq. (\ref{eqn:sus})).\\

In Fig. \ref{fig:U4st} the GBC is shown as a function of reduced energy for different values of $W$.  The system sizes considered were $L=100, 200, 400$, the number of disorder realizations over which averages were taken was $1000$.  The reduced energy range was divided into one hundred equal segments.  Averages were taken within each segment.  The Binder cumulants are smaller at the band edges, and they reach their maxima around the band center.  In the interval $0<W<1$ the system is known to be localized~\cite{Abrahams79}, but this is not easy to establish numerically.  The GBC is often greater than zero, especially for $W=0.1$, but  its value never reaches $0.5$ In the interval $0<W<1$ for the system sizes studied.  Also, in all three cases ($W=0.1, 0.5, 1.0$) the tendency of the GBC when system size is increased is to decrease, for example, for $W=1.0$, the largest system size is below zero for the entire band.  The fact that $U_4>0$ for $W=0.1$ and near the band center for $W=0.5$ is due to the fact that the correlation length is comparable to the system size for these cases.   Essentially the same conclusions can be drawn for the many-body case as a function of particle density, $\rho$, (Fig. \ref{fig:U4sd_1D}).  \\

\begin{figure}[t]
 \centering
\includegraphics[width=8cm,keepaspectratio=true]{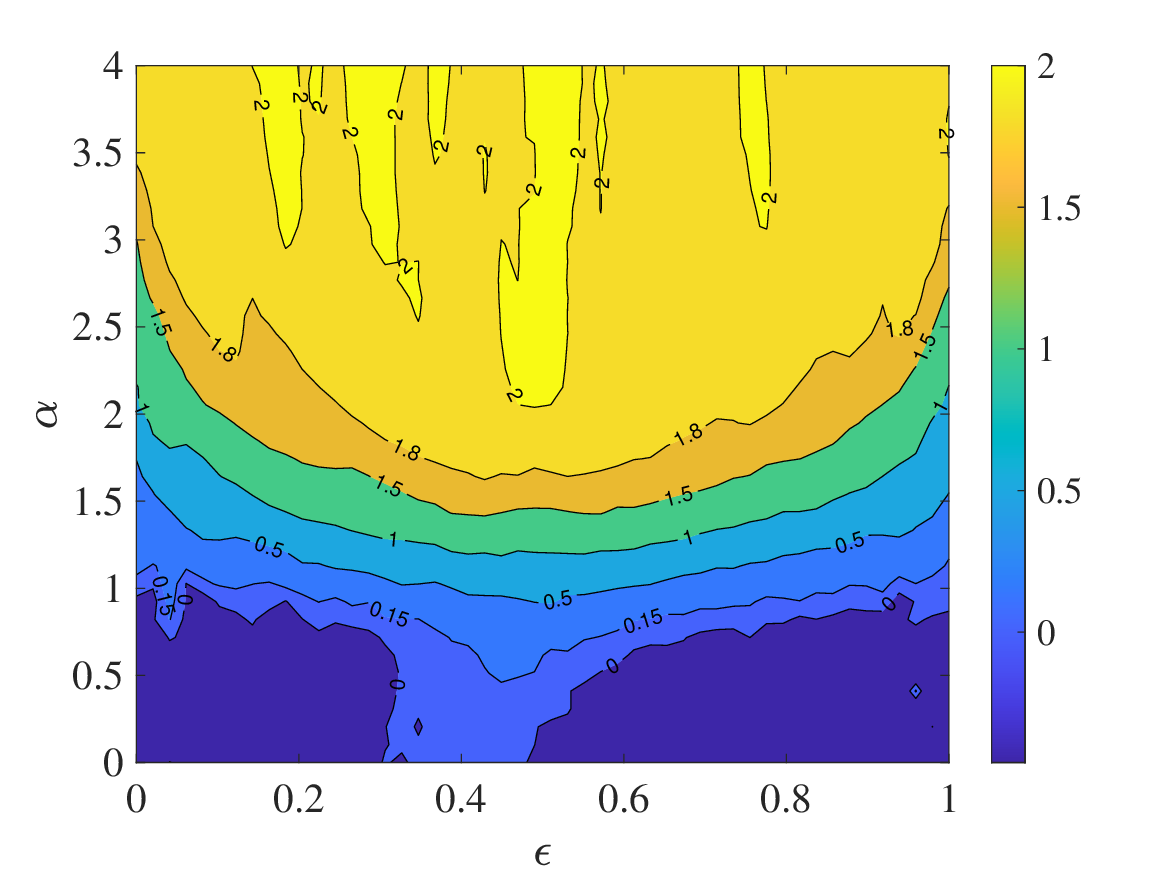}
 \caption{Contour plot of the size scaling exponent, $\gamma$ as a function of reduced energy, $\epsilon$, and $\alpha$, determined from a fit to the following system sizes ($L=512, 1024, 2048, 4096$) of the de Moura-Lyra model.  The number of disorder realizations was $480, 240,120, 60$, respectively.}
 \label{fig:gamma_dML}
\end{figure}

\begin{figure}[t]
 \centering
\includegraphics[width=8cm,keepaspectratio=true]{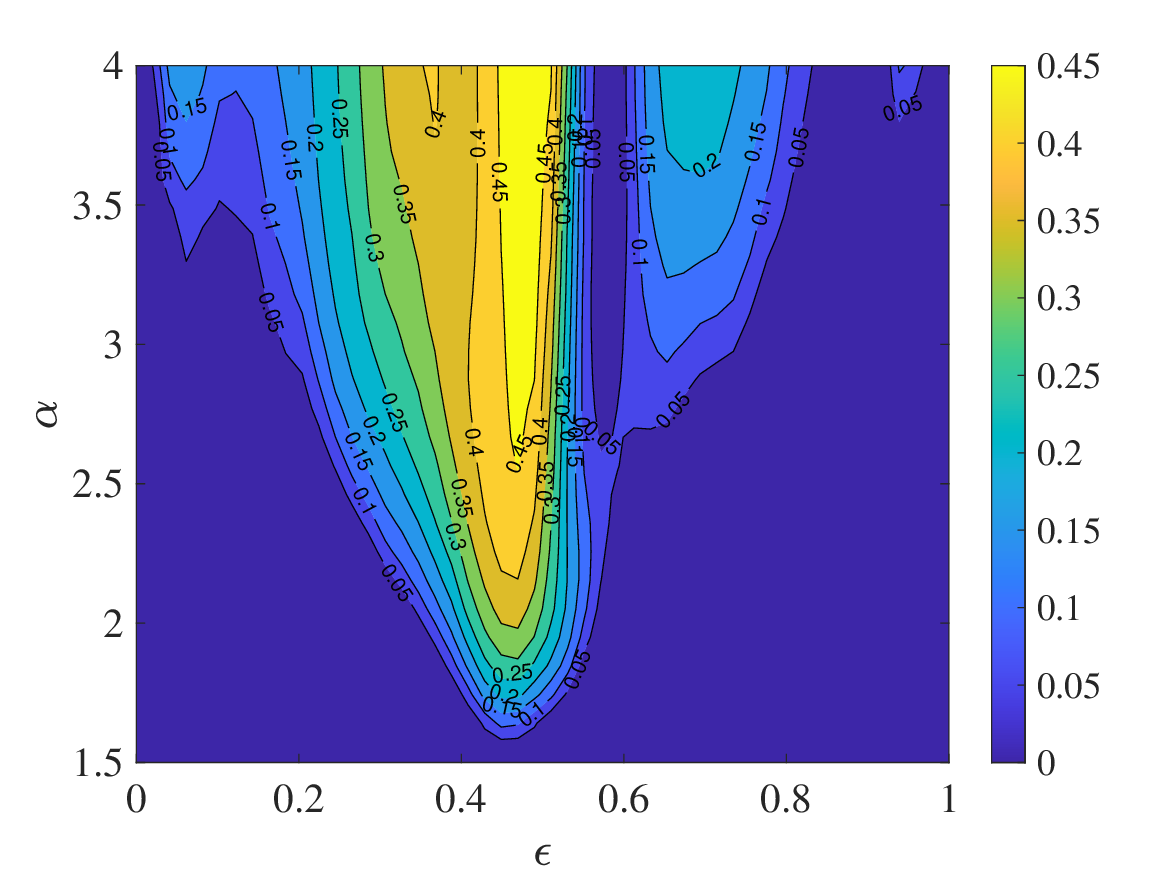}
\includegraphics[width=8cm,keepaspectratio=true]{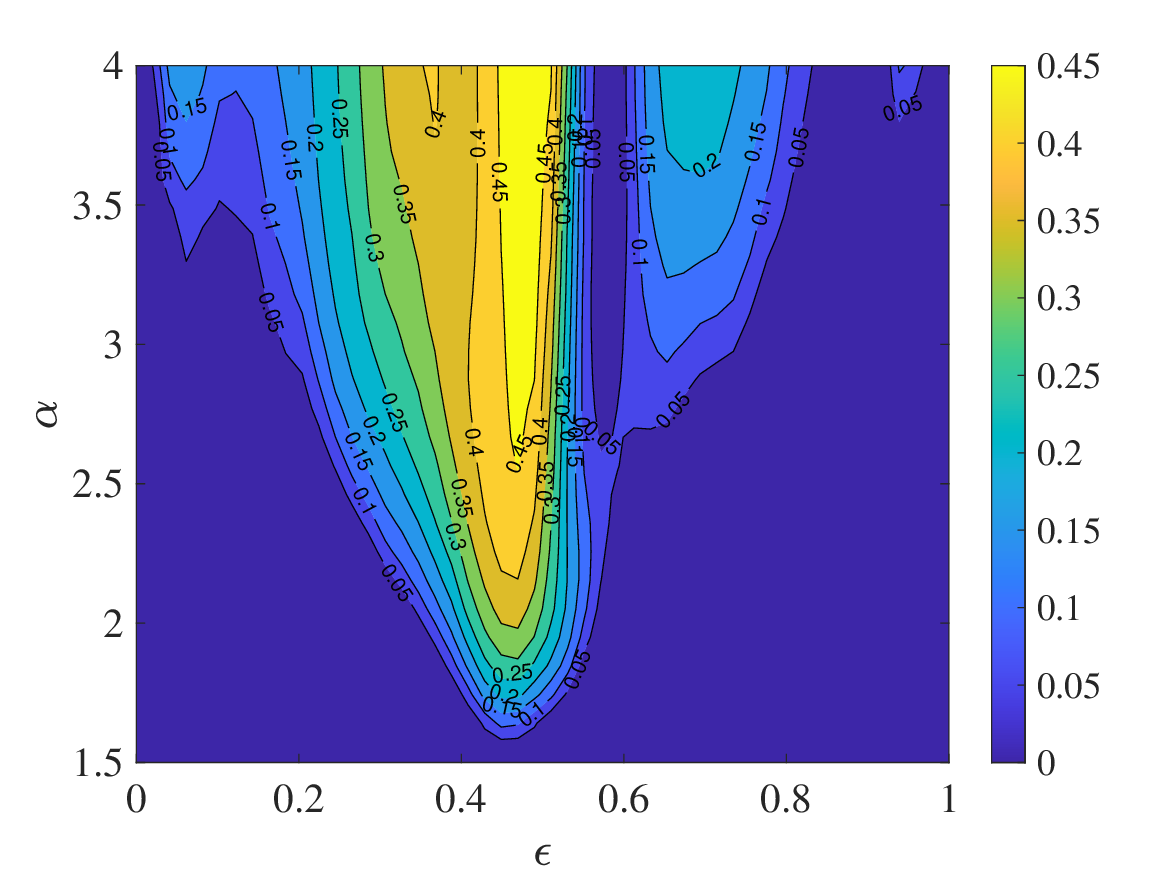}
\includegraphics[width=8cm,keepaspectratio=true]{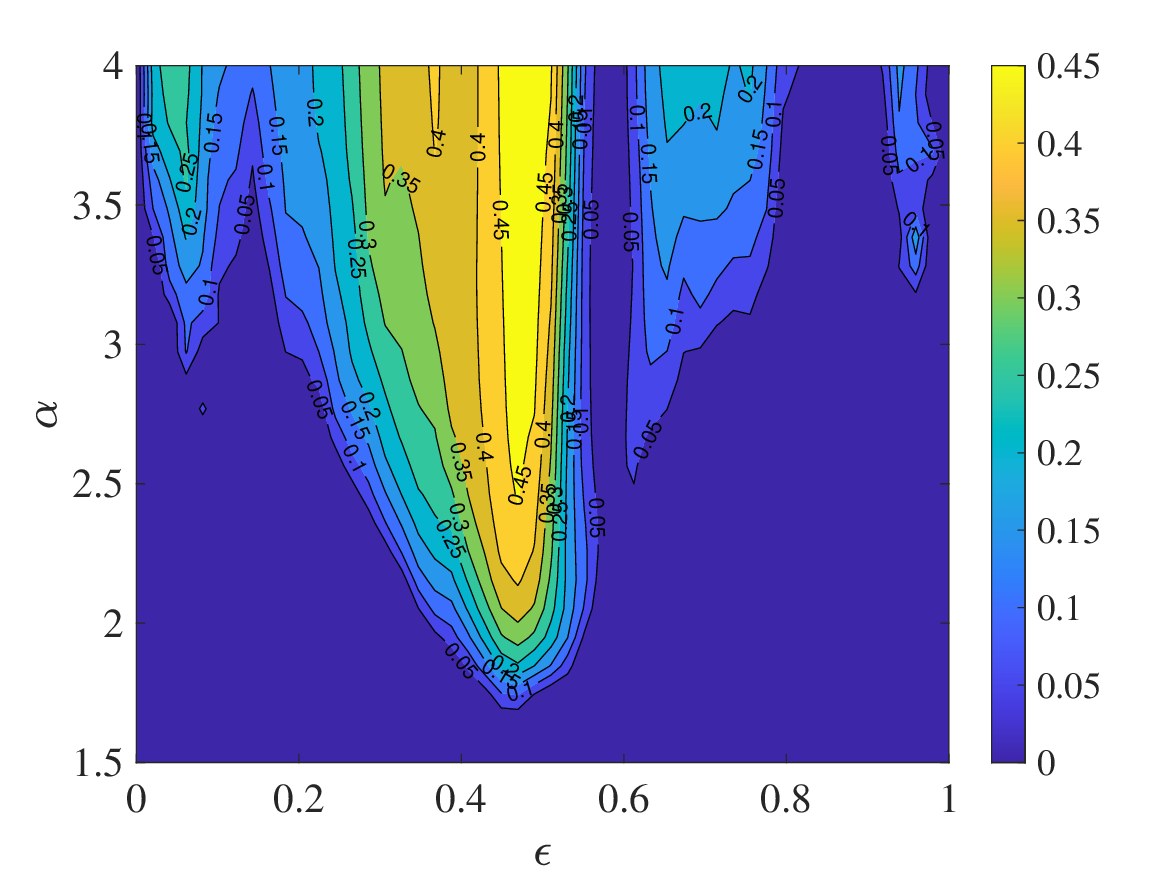}
 \caption{Geometric Binder cumulant, $U_4$, as a function of the reduced energy, $\epsilon$, and $\alpha$ for systems of size $L=1024, 2048, 4096$.  The number of disorder realizations was $240$, $120$, and $60$, respectively.  At points where $U_4<0$, $U_4$ was set to zero.}
 \label{fig:U4_dML}
\end{figure}

\begin{figure}[t]
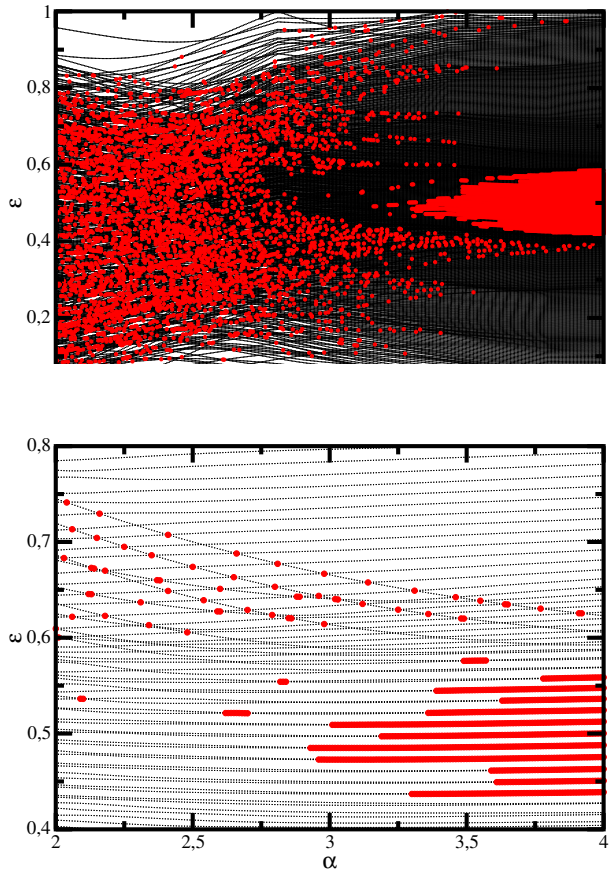

 \centering
\includegraphics[width=8cm,keepaspectratio=true]{./nrgL512.eps}
\includegraphics[width=8cm,keepaspectratio=true]{./nrgL128.eps}
 \caption{Upper panel: analysis of the energy spectrum as a function of $\alpha$ for a system of size $L=512$.  The black plus signs are all energy eigenvalues for a single disorder realization as they evolve with changing $\alpha$.  The red filled circles show energy levels for which the next highest level is within a certain threshold, $.05/L$.  For $\alpha<2$ red circles are found randomly, at level crossings.  For $\alpha>2$, in the reduced energy range $\approx 0.4 - 0.6$ actual gap closures occur which persist for finite intervals of $\alpha$.  Lower panel: a zoomed version of the previous panel for a system of $L=128$.   }
 \label{fig:nrg}
\end{figure}

\section{Results: de Moura-Lyra model}

\label{sec:dML}

The de Moura-Lyra model was proposed to account for power law correlated disorder.  The disorder potential (on-site energies on lattice sites) is constructed via an inverse Fourier transform of a function with $k$-vectors whose power law dependence is controlled by a parameter $\alpha$.  The model exhibits a number of pathologies.   The interest in the model stemmed in large part from the results of early studies which indicated the presence of a mobility edge.  de Moura and Lyra calculated~\cite{deMoura98} the Lyapunov exponent which showed that states near the band center for $\alpha>2$ are extended.   This conclusion was confirmed by a study~\cite{Dominguez-Adame03} of Bloch oscillations.  However, on this point, some studies started showing contradictory results~\cite{Shima04,Kaya07}.  Petersen and Sandler~\cite{Petersen13} analyzed the correlation function of the on-site energies and showed that for $\alpha>1$ a size dependent anti-correlation is present.  For large $\alpha$ the disorder potential was shown to be a sinusoidal superpotential when viewed at large length scales with randomness still persisting at small lengthscales.   The mobility edge was explicitly questioned by the work of Santos Pires {\it et al.}\cite{SantosPires19} .   In this study the correlation of on site energies was considered.  It was shown that a short range uncorrelation exists whose presence persists to large system sizes.  Replacing its contribution with its value in the thermodynamic limit leads to a simple scaling function based on which the thermodynamic limit of the localization length can be obtained from extrapolation.  These and other numerical results show a global delocalization transition in the dMLM at $\alpha = 1$, and since, for $\alpha>1$ all states are delocalized, there can be no mobility edge.\\

The Hamiltonian of the dMLM is given by
\begin{equation}
\label{eqn:H_dML}
H  =  \sum_{j=1}^L \left[ -t(c_j^\dagger c_{j+1} + c_{j+1}^\dagger c_j) + W \xi_j n_j \right],
\end{equation}
 where $c_j^\dagger$($c_j$) denote the creation(annihilation) operators at site $j$, $n_j = c_j^\dagger c_j$ denotes the particle density operator at site $j$, $t$ denotes the hopping parameter and $W$ denotes the disorder strength, which is taken to be unity.   $\xi_j$ denote the on-site valued of the disorder potential.\\

 The disorder potential in this case is generated as follows.  First a set of random numbers $\phi_k \in [0,2\pi]$ are generated.  From $\phi_k$ $\xi_j$ are defined as
 \begin{equation}
 \xi_j = \sum_{k=1}^{\frac{L}{2}} \left[ \left( \frac{2\pi}{L} \right)^{1-\alpha} \frac{1}{k^\alpha}\right]^{\frac{1}{2}} \cos \left( \frac{2 \pi j k}{L} + \phi_k \right),
 \end{equation}
 but before substituting in Eq. (\ref{eqn:H_dML}), $\xi_j$ are shifted so that their average is zero, and scaled so that their root mean square average is unity. \\
  
 \begin{figure}[t]
 \centering
\includegraphics[width=8cm,keepaspectratio=true]{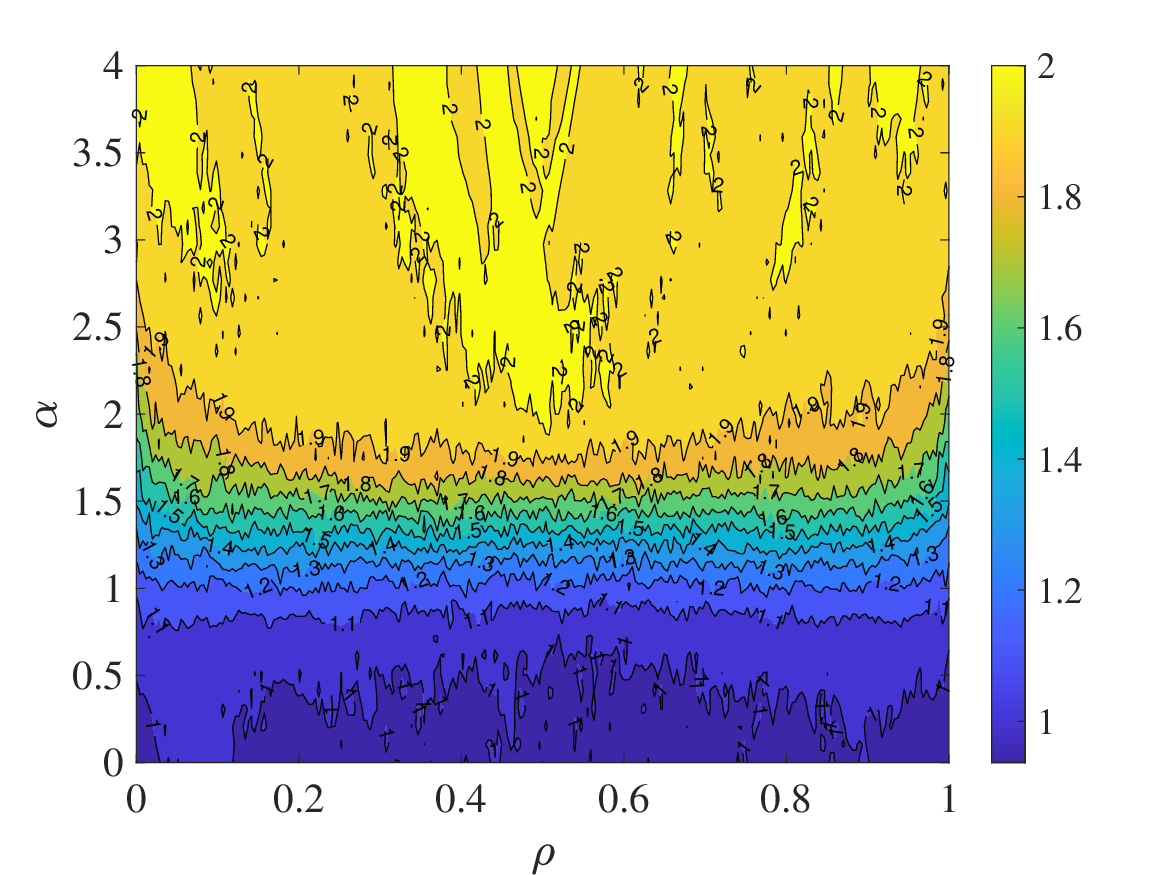}
 \caption{Contour plot of the size scaling exponent, $\gamma$ as a function of particle density, $\rho$, and $\alpha$, determined from a fit to the following system sizes ($L=512,1024,2048,4096$) of the de Moura-Lyra model.  The number of disorder realizations was $480, 240, 120, 60$, respectively.}
 \label{fig:gamma_dML_SD}
\end{figure}
 
 In Fig. \ref{fig:gamma_dML} the size scaling exponent is presented as a function of reduced energy, $\epsilon$, and $\alpha$.    Four system sizes were used, $L=512,1024, 2048, 4096$, the number of disorder realizations averaged were, $480, 240, 120, 60$, respectively.  $\gamma$ is near zero at the band edges and at small $\alpha$.  Even at $\alpha=0$ $\gamma$ increases near the band center, but it remains around $\gamma \approx 1$ for $\alpha<1$.  For $\alpha>1$ $\gamma$ starts a rapid increase, reaching $\gamma \approx 2$ by $\alpha \approx 2$.   The presence of a mobility edge is questionable from these results, since the increase in $\gamma$ occurse for all $\epsilon$.   These results are consistent with the findings of Santos Pires {\it et al.}~\cite{SantosPires19}: the value of $\gamma$ suggests delocalization for $\alpha > 1$. \\
 
 In Fig. \ref{fig:U4_dML} the GBC ($U_4$), is shown as a function of reduced energy, $\epsilon$, and the parameter $\alpha$.  Three system sizes are shown, $L=1048, 2048, 4096$, the number of disorder realizations averaged were $240, 120, 60$, respectively.   $U_4$ also tends to increase as a function of $\alpha$, but here this does not occur for all values of $\epsilon$.  The delocalization, measured by the GBC, rather than only the scaling exponent, is most intense near the band center.   A more complete understanding is gained by investigating the energy levels and their spacings, shown in Fig. \ref{fig:nrg}.    This graph shows how energy eigenvalues evolve as a function of $\alpha$ for two system sizes (upper panel, $L=512$, lower panel $L=128$).  The black lines correspond to such energy states.  The red dots indicate states $\epsilon_n$ such that $\epsilon_{n+1}-\epsilon_n$ is below the threshold values $0.05/L$.    Red dots are found for $\alpha<2$, because isolated level crossings do occur.  What is interesting is that for the region, $\alpha > 2.5$, $0.4 < \epsilon < 0.6$ the red dots appear to form lines.  This is seen even better for the smaller system size, $L=128$ (lower panel).  The states in the band center tend to close gaps, which remain closed after further increase in $\alpha$.  This does not happen for $\alpha<2$, nor for larger values of $\alpha$ around the band edges.   Most crucially, we identified a tendency in the region that was thought by early studies of this model to be a mobility edge: pairs of degenerate states form.\\
 
 \begin{figure}[t]
 \centering
\includegraphics[width=8cm,keepaspectratio=true]{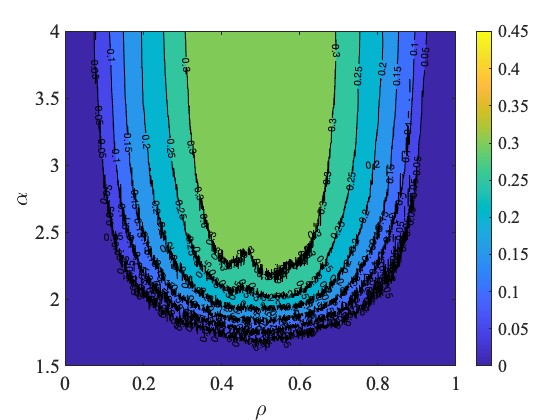}
\includegraphics[width=8cm,keepaspectratio=true]{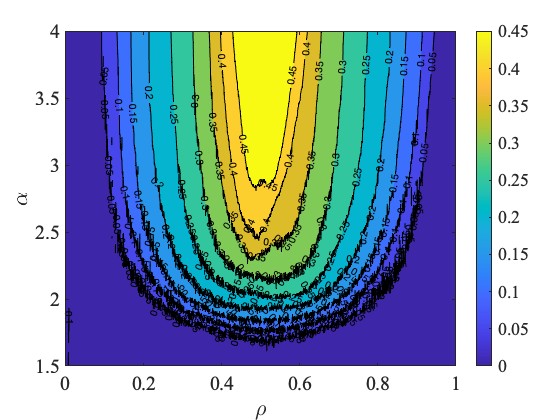}
 \caption{Contour plot of the geometric Binder cumulant $U_4$ for the de Moura-Lyra model as a function of $\alpha$ and $\rho$, particle density for a system size of $L=4096$.  The number of disorder samples averaged was $60$.  The upper panel shows for even particle, number, the lower one for odd particle number. } 
 \label{fig:U4_dML_SD}
\end{figure}

\begin{figure}[t]
 \centering
\includegraphics[width=8cm,keepaspectratio=true]{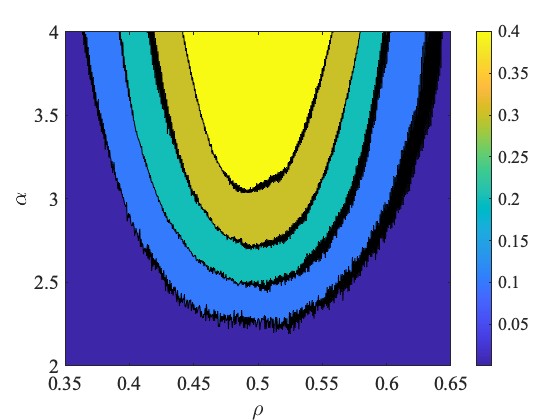}
 \caption{Contour plot of the degeneracy indicator for the de Moura-Lyra model.  The system was of size $L=2048$ and the number of disorder realizations was $120$. }
 \label{fig:zdiff}
\end{figure}

\begin{figure}[t]
 \centering
\includegraphics[width=8cm,keepaspectratio=true]{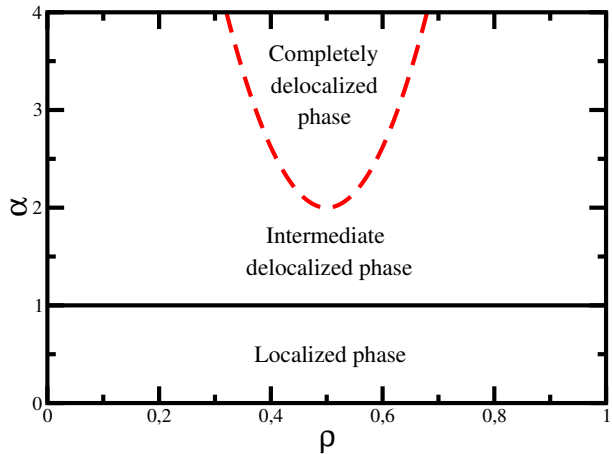}
 \caption{Phase diagram of the de Moura-Lyra model.}
 \label{fig:pd_dML}
\end{figure}

Our calculations on a many-body system concur with the above findings, but, interestingly, the calculations are more stable.  The scaling exponent, $\gamma$,  shows similar behavior to the single level case (Fig. \ref{fig:gamma_dML_SD}), except that it floors at a value of $\gamma \approx 1$, rather than $\gamma \approx 0$ (which was also found for the Anderson model).    At $\alpha \approx 1$, $\gamma$ grows and monotonically increases to $\gamma \approx 2$.  The tendency to delocalize appears somewhat weaker at the band edges, but it happens for all values of the density, $\rho$.   Again, these results are consistent with a delocalization transition at $\alpha \approx 1$, as stated by Santos Pires {\it et al.}~\cite{SantosPires19}.  The GBC results, shown in Fig. \ref{fig:U4_dML_SD} for the system size $L=4096$ ($60$ disorder realizations), indicate strong localization around $\rho \approx 0.5$, and for $\alpha >2$, if the number of particles is odd.  The maximum $U_4 \approx 0.5$ and this maximum is reached near the band center, for $\alpha>2$.   Localization is also found for even particle number, but less pronounced ($U_4 \approx 0.3$ at its maximum value).   This is expected from the energy gap closures found in Fig. \ref{fig:nrg}.  The degeneracy indicator is shown in Fig. \ref{fig:zdiff}.   Again, there is a definite region, $\alpha > 2$, around a filling density of $\rho \approx 0.5$, where a degeneracy is seen.  Both the GBC and the degeneracy indicator can be considered as gauges of localization, reaching their maximum values at maximal delocalization.   We summarize our results by proposing a phase diagram for the dMLM as a function of $\alpha$ and particle density, $\rho$, in  Fig. \ref{fig:pd_dML}. \\

 \section{Conclusion}
 
 \label{sec:cnclsn}

 We reported disorder averaged calculations for the localization sensitive quantities of the modern theory of polarization for two different one dimensional models.  Our theoretical/methodical extensions render the modern theory of polarization quantitatively applicable to disordered systems offering a number of quantities sensitive to the details of localization, such as the variance of the polarization, the geometric Binder cumulant, the size scaling exponent, and an indicator based on the degeneracy of the wave function induced by the application of a Peierls substitution.  The simultaneous use of these quantities gives a complete picture of localization in disordered systems.  Our example calculations show this for the simple one dimensional Anderson model, as well as for the de Moura-Lyra model.\\
 
Our results refine the existing picture regarding the mobility edge in the de Moura-Lyra model  As the correlation parameter for the power law decay, $\alpha$, increases, delocalization sets in at a finite value, separating localized from delocalized states.  However, we showed that there is another phase line which separates the delocalized phase into two distinct regions.  The region near the band center for which $\alpha>2$ is characterized by persistent gap closures between pairs of states as a function of $\alpha$.  In the many-body case, when the model is filled with fermions, this gives rise to an alternation between even and odd particles in the delocalization sensitive quantities.  They reach their maximum values only for odd particle number, which corresponds to half filled conduction bands.\\

\section*{Acknowledgments}    BH gratefully acknowledges support by HUN-REN 3410107 (HUN-REN-BME-BCE Quantum Technology Research Group), by the National Research, Development and Innovation Fund of Hungary within the Quantum Technology National Excellence Program (Project No. 2017-1.2.1-NKP-2017-00001), by Grants No. K142179, No. K142652, and No. FK142601 and by the BME-Nanotechnology FIKP Grant No. (BME FIKP-NAT).  LMM gratefully acknowledges FCT-Portugal through  Grant No. UID/04650 -  CFUMUP, Centro de Física das Universidades do Minho e do Porto, Portugal.  AL gratefully acknowledge financial support by the National Research, Development, and Innovation Office (NRDI) of Hungary under Project Nos. FK142601, K142652. BH and AL acknowledge the Digital Government Development and Project Management Ltd. for awarding us access to the Komondor HPC facility based in Hungary.\\


\begin{thebibliography}{9}
\bibitem{MBLnote} The term many-body localization, nowadays, refers to the ergodicity breaking that occurs in certain many-body systems.  In this particular sentence, however, we are referring to the localization of the center of charge of a crystalline system.
\end{thebibliography}
\end{document}